\documentclass[prd,twocolumn,showpacs,preprintnumbers,amsmath,amssymb,superscriptaddress,floatfix,nofootinbib]{revtex4}
\usepackage{graphicx}
\usepackage{bm}
\usepackage{amsmath}  % 用于数学符号
\usepackage{amsfonts}
\usepackage{amssymb}
\usepackage{color}
\usepackage{multirow}
\usepackage{subfigure}
\usepackage{booktabs} % 用于绘制专业表格线
\usepackage{overpic} % 叠加文字到图上

\usepackage[colorlinks, citecolor=blue,anchorcolor=red,menucolor=red, linkcolor=red,filecolor=red,runcolor=red,urlcolor=blue,frenchlinks=red]{hyperref}

\allowdisplaybreaks[1]
\begin{document}

\title{Roles of $f_0(980)$ and $a_0(980)$ in the process $\Lambda_c^+ \to pK_S^0K_S^0$: a clean probe of the $K\bar K$ final-state interaction}

\author{Li-Na Zhou}
\affiliation{School of Physics, Zhengzhou
	University, Zhengzhou 450001, China}

\author{Le-Le Wei}\email{lelewei@zzu.edu.cn}
\affiliation{School of Physics, Zhengzhou
	University, Zhengzhou 450001, China}

    \author{Li-Juan Liu}\email{liulijuan@zzu.edu.cn}
\affiliation{School of Physics, Zhengzhou
	University, Zhengzhou 450001, China}

\author{En Wang}\email{wangen@zzu.edu.cn}
\affiliation{School of Physics, Zhengzhou
	University, Zhengzhou 450001, China}

\begin{abstract}

Based on the Belle measurement of $\Lambda_c^+ \to pK_S^0K_S^0$, we study the contributions of the light scalar mesons $f_0(980)$ and $a_0(980)$, dynamically
generated from the $S$-wave pseudoscalar meson--pseudoscalar meson interaction within the chiral unitary approach, to the threshold enhancement in the
$K_S^0K_S^0$ invariant mass squared distribution, and examine the possible contribution of the intermediate $\Sigma(1620)$ resonance in the $pK_S^0$ subsystem.
We first argue that the $\Lambda_c^+ \to pK_S^0K_S^0$ channel is a cleaner probe of the $K\bar K$ final-state interaction than the charged channel
$\Lambda_c^+ \to pK^+K^-$, since the $\phi$ and $\Lambda^*$ contributions are forbidden and the $\Sigma(1620)$ and $\Sigma(1660)$ couple weakly to the $\bar{K}N$ channel. 
The threshold enhancement in the $K_S^0K_S^0$ invariant mass squared distribution is well described by the $S$-wave $K^0\bar K^0$ final-state interaction alone,
with the $f_0(980)$ contribution dominating while that of $a_0(980)$ is small, supporting the interpretation of $f_0(980)$ as a $K\bar K$ hadronic
molecular state. Future high-precision
measurements by Belle~II and the proposed Super Tau-Charm Facility will help test the roles of the $f_0(980)$ and $a_0(980)$ and of the possible
$\Sigma(1620)$ contribution in this process.

%Based on the experimental measurement of the Belle Collaboration for the $\Lambda_c^+ \to p K_S^0 K_S^0$ decay, within the framework of the chiral unitary approach, this work investigates the contributions of the hadronic molecular states $f_0(980)$ and $a_0(980)$, dynamically generated from the $S$-wave pseudoscalar meson--pseudoscalar meson interaction, to the threshold enhancement structure in the $K_S^0 K_S^0$ invariant mass distribution. Meanwhile, the possible contributions of the intermediate resonance $\Sigma(1620)$ in the $p\bar{K}^0$ subsystem are also examined. The fit indicates that the threshold enhancement is well described by only the $S$-wave final-state interaction of $K^0\bar{K}^0$ alone. The $f_0(980)$ contribution is dominant, whereas that of $a_0(980)$ is relatively small. This result supports the interpretation of $f_0(980)$ as a hadronic molecular state. Including $\Sigma(1620)$ does not significantly improve the fit, and the available data are insufficient to determine their contributions. Future measurements with higher precision will help test the roles of these resonances in this process.

\end{abstract}

\maketitle

%%%%%%%%%%%%%%%%%%%%%%
\section{Introduction} \label{sec:Introduction}

In the light-quark sector, the light scalar-meson spectrum below 1~GeV has long
been a subject of debate in hadron physics~\cite{Close:2002zu,Pelaez:2015qba}.
The $f_0(980)$ and $a_0(980)$ are light scalar mesons with masses near 980~MeV
and $J^P=0^+$, with isospins $I=0$ and $I=1$, respectively. The $f_0(980)$
predominantly couples to the $\pi\pi$ and $K\bar K$ channels, whereas the
$a_0(980)$ predominantly couples to the $\pi\eta$ and $K\bar K$
channels~\cite{OllerOsetPelaez1999,Wang2021a0f0,Abreu2023}. Since both states
lie near the $K\bar K$ threshold and couple strongly to the $K\bar K$ channel,
their properties are closely connected to the $K\bar K$ two-body
interaction~\cite{Abreu2023}. Their masses and strong couplings are difficult
to accommodate within a simple $q\bar q$ meson picture alone. Several
interpretations have been proposed for the internal structures of the
$f_0(980)$ and $a_0(980)$, including conventional $q\bar q$
states~\cite{Maiani:2004uc}, compact tetraquark states~\cite{Maiani:2004uc},
$K\bar K$ hadronic molecular states~\cite{Guo:2017jvc}, and mixed states with
multiple components~\cite{Wolkanowski:2015lsa}.

Nonleptonic weak decays of the charmed baryon $\Lambda_c^+$ offer a large phase
space and involve complex final-state strong interactions, making them an ideal
setting for investigating light scalar mesons and various
resonances~\cite{Lai:2025ryi,Li:2025gvo,Li:2026wbl,Li:2026umb,Li:2026lbo,Zhang:2026igc,Li:2025exm}. In a
previous study, Wang \textit{et al.}~\cite{Wang:2020pem} investigated the
singly Cabibbo-suppressed decay $\Lambda_c^+ \to pK^+K^-$ within the chiral
unitary approach~\cite{OllerOsetPelaez1999,Wang2021a0f0}, including
the $S$-wave $K^+K^-$ final-state interaction and the contribution from the
intermediate vector meson $\phi$. The theoretical results agree well with the
BESIII data~\cite{BESIII:2016ozn}. The $K^+K^-$ invariant mass squared distribution
exhibits not only the narrow peak associated with the $\phi$ meson but also a
pronounced enhancement near the $K^+K^-$ threshold, the latter being attributed
to the $S$-wave $K^+K^-$ final-state interaction that dynamically generates the
$f_0(980)$ and $a_0(980)$
resonances~\cite{OllerOsetPelaez1999,Guo:2017jvc,Wang:2020pem}. This
demonstrates that the $K\bar K$ invariant mass squared distribution in three-body
$\Lambda_c^+$ decays can be used to study the dynamical properties of the
$K\bar K$ final-state interaction and the associated light scalar resonances,
providing a theoretical reference for the $K\bar K$ system in three-body
$\Lambda_c^+$ decays~\cite{Wang2021a0f0,Abreu2023,Wang:2022nac,Hyodo:2011js,
Zhang:2024jby}.

The role of the $I=0$ and $I=1$ components in $K\bar K$-related three-body
decays has by now been established in several different settings. In the
charmless $B^- \to \pi^-K^+K^-$ and $B^- \to \pi^-K^0\bar K^0$ reactions, an
analysis within the chiral unitary SU(3) formalism shows that the $I=0$
component associated with the $f_0(980)$ generates the dominant contribution in
the low $K^+K^-$ invariant mass region, while the $I=1$ component tied to the
$a_0(980)$ is comparatively small~\cite{Abreu2023}. The same qualitative
pattern is expected to persist in charmed-baryon decays, and one of the goals
of the present work is to test this expectation in a channel where the $K\bar
K$ dynamics can be isolated more cleanly.

It is also worth noting that the light pseudoscalar sector around 1.4~GeV, in
particular the $\eta(1405)/\eta(1475)$ system, has attracted considerable
attention in the radiative charmonium decays $J/\psi \to \gamma
K_S^0K_S^0\pi^0$, where a partial wave analysis of the $K_S^0K_S^0\pi^0$
system with the covariant tensor amplitude method identifies dominant
pseudoscalar and axial-vector
components~\cite{BESIII:2022chl}. This example illustrates both the power and
the subtlety of amplitude analyses in $K\bar K$-containing final states: even
with a large data sample, disentangling overlapping structures in the
$K\bar K(\pi)$ system requires a consistent treatment of the dynamics. For the
$\Lambda_c^+ \to pK_S^0K_S^0$ channel, the situation is simplified because the
relevant light scalars $f_0(980)$ and $a_0(980)$ appear through the $S$-wave
$K\bar K$ final-state interaction, and their interpretation as dynamically
generated states can be tested directly.

In 2022, the Belle Collaboration reported the first measurement of the
branching fraction for $\Lambda_c^+ \to pK_S^0K_S^0$, obtaining
$\mathcal{B}(\Lambda_c^+ \to pK_S^0K_S^0) = (2.35 \pm 0.12 \pm 0.07 \pm 0.12)
\times 10^{-4}$ with a statistical significance exceeding
$10\sigma$~\cite{Belle:2022pwd}, together with the $K_S^0K_S^0$ and $pK_S^0$
invariant mass squared distributions. Belle observed a significant enhancement
near the $K_S^0K_S^0$ threshold and suggested that it may be associated with
$f_0(980)$ or $a_0(980)$. The same measurement also reported the channel
$\Lambda_c^+ \to pK_S^0\eta$, in which the final-state $p\eta$ is in a pure
$I=1/2$ state and is therefore suited to studying the $N^*(1535)$ resonance;
the two light scalars $a_0(980)$ and $f_0(980)$ were noted as possible
intermediate states in the $K\bar K$ subsystem of the $pK_S^0K_S^0$
mode~\cite{Belle:2022pwd}. The fact that the Belle analysis could not separate
the $I=0$ and $I=1$ contributions from the $K_S^0K_S^0$ spectrum alone
motivates a dedicated theoretical treatment in which both isospin channels are
described consistently.

We therefore stress the following key point of the present work: the
$\Lambda_c^+ \to pK_S^0K_S^0$ channel is a cleaner laboratory for the $K\bar K$
final-state interaction than the charged channel $\Lambda_c^+ \to pK^+K^-$.

\emph{First, the $\phi(1020)$ contamination is removed by symmetry.} The $\phi$
meson produces a $K^0\bar K^0$ pair in a $P$-wave (antisymmetric) state, whereas
$K_S^0K_S^0$ is a symmetric state of two identical bosons; the projection of the
$\phi$-mediated amplitude onto $K_S^0K_S^0$ therefore vanishes. As a result, the
narrow $\phi$ peak that dominates the $K^+K^-$ spectrum is absent here and
cannot mask the near-threshold structure.

\emph{Second, no $\Lambda$ excitation can contribute.} The two-body subsystems
$pK^0$ and $p\bar K^0$ both carry charge $+1$, while the $\Lambda$ excitations
[$\Lambda(1405)$, $\Lambda(1670)$, etc.] are neutral; by charge conservation
none of them can decay into either subsystem, and hence the $pK_S^0$ spectrum
is free of $\Lambda$-excitation background. The only possible baryon-resonance
background is a $\Sigma$-type excitation. These features make this channel
particularly suitable for isolating the $K\bar K$ final-state interaction and
for disentangling the $I=0$ $f_0(980)$ and the $I=1$ $a_0(980)$.

Because the $K_S^0K_S^0$ final state contains both $I=0$ and $I=1$ components,
and the weak decay of $\Lambda_c^+$ does not conserve isospin, both the $I=0$
$f_0(980)$ and the $I=1$ $a_0(980)$ can couple to it through the $S$-wave
final-state
interaction~\cite{Wang:2022nac,Li:2024rqb,Li:2025gvo,Wang2021a0f0}. Thus the
two states cannot be distinguished solely from the $K_S^0K_S^0$ invariant mass
squared distribution, and their relative weight must be analysed within a consistent
coupled-channel treatment.

Before turning to the dynamics, we recall that the description of these two
states often relies on the Flatt\'e parametrization, which retains only the
coupling to the $K\bar K$ channel and neglects the other channels; for the
$I=1$ state $a_0(980)$ this excludes the $\pi\eta$ channel, which is open and
lies close to the $K\bar K$ threshold. Such a truncation distorts the extracted
line shape and parameters, and through the $f_0(980)$--$a_0(980)$ interference
it also affects the $K_S^0K_S^0$ spectrum. The chiral unitary approach adopted
here avoids this truncation by treating all relevant channels on the same
footing.

Belle also reported the $pK_S^0$ invariant mass squared distribution of the process $\Lambda_c^+ \to pK_S^0K_S^0$~\cite{Belle:2022pwd}, which can
be used to investigate possible intermediate baryon
resonances~\cite{Miyahara:2015cja}. Mild
excesses relative to phase space are observed around $M_{pK_S^0}^2 \approx
2.6$--$2.8~\mathrm{GeV}^2$, close to the masses of low-lying strange baryon
states such as $\Sigma(1620)$. The Belle
analysis focused primarily on the near-threshold structure in the
$K_S^0K_S^0$ spectrum and did not assign a specific baryon-resonance
interpretation to the $pK_S^0$ distribution. This motivates examining whether a
$\Sigma(1620)$ contribution is required by the present
data.

Finally, the study of three-body charmed-baryon decays such as $\Lambda_c^+ \to
pK_S^0K_S^0$ is in itself of broad interest, as these processes provide access
to the spin structure and polarization properties of the decay, and can be used
to test the flavor symmetry of the underlying weak
Hamiltonian~\cite{Cen:2019ims}. A reliable description of the strong-interaction
dynamics in the $K\bar K$ subsystem therefore has implications beyond the
spectroscopy of the light scalars themselves.

Motivated by these considerations, we investigate the $\Lambda_c^+ \to
pK^0\bar K^0$ decay within the chiral unitary approach, based on the Belle
measurement of $\Lambda_c^+ \to pK_S^0K_S^0$. We first study the near-threshold
enhancement in the $K_S^0K_S^0$ invariant mass squared distribution by considering the
$S$-wave $K^0\bar K^0$ final-state interaction and by separating the $I=0$ and
$I=1$ contributions, so as to examine the respective roles of the dynamically
generated $f_0(980)$ and $a_0(980)$. We then use the $pK_S^0$ invariant mass squared
distribution to test a possible intermediate $\Sigma(1620)$ contribution in the
$p\bar K^0$ subsystem. We stress that the chiral-unitary amplitudes used here,
which incorporate the full coupled-channel dynamics~\cite{Lyu:2026ack}, provide a more faithful
description of the $f_0(980)$ and $a_0(980)$ line shapes near the $K\bar K$
threshold than a Flatt\'e parametrization restricted to the $K\bar K$ channel
alone. This work thus clarifies the role of final-state interactions in the
three-body weak decay of $\Lambda_c^+$ and sharpens our understanding of the
hadronic structure of $f_0(980)$ and $a_0(980)$.

The structure of this paper is as follows. The theoretical framework is
presented in Sec.~\ref{sec:Formalism}. Results and discussions are presented in
Sec.~\ref{sec:Results and Discussion}. Finally, a summary is given in Sec.~\ref{sec:summary}.

\section{Formalism} \label{sec:Formalism}

In this section, we present the theoretical framework for the decay process
$\Lambda_c^+ \to pK_S^0 K_S^0$. In Sec.~\ref{sec2a} we describe the mechanism
for the dynamical generation of the $f_0(980)$ and $a_0(980)$ states through the
$S$-wave final-state interaction in the $K^0\bar K^0$ system. In
Sec.~\ref{sec2b} we consider the possible contribution of the intermediate
baryon resonance $\Sigma(1620)$ in the $p\bar K^0$ subsystem. Finally, in
Sec.~\ref{sec2c} we present the theoretical expressions for the
invariant mass squared distributions.

\subsection{$S$-wave meson-meson interaction and $f_0(980)/a_0(980)$}\label{sec2a}

%图1
\begin{figure}[tbhp]
	\centering
	\includegraphics[scale=0.9]{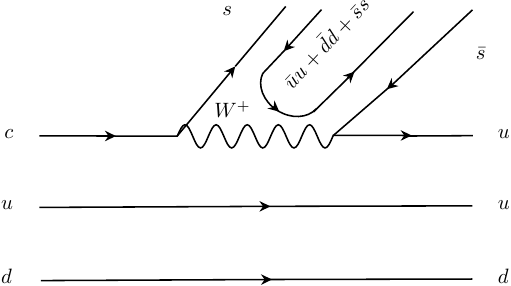}
	%\vspace{-0.7cm}
	\caption{Quark-level diagram of the $W^+$ internal-emission mechanism for the $\Lambda_c^+\to pK^0\bar K^0$ decay.}
	\label{fig:internalW}
\end{figure}

%图1
%\begin{figure}[tb]
%\centering
%\includegraphics[width=0.78\columnwidth]%{figures/fig1_internal_w.png}
%\caption{Quark-level diagram of the $W^+$ internal-emission mechanism for the $\Lambda_c^+\to pK^0\bar K^0$ decay.}
%\label{fig:internalW}
%\end{figure}

Following Refs.~\cite{OllerOsetPelaez1999,Wang:2020pem,Wang:2022nac}, we adopt the internal $W$-emission mechanism to describe the weak decay process
$\Lambda_c^+ \to pK^0\bar K^0$, as shown in Fig.~\ref{fig:internalW}. In the weak decay of the $\Lambda_c^+$, the $c$ quark decays into a $W^+$ boson and an $s$ quark, while the $W^+$ subsequently decays into an $\bar s u$ pair. To produce the final state $pK^0\bar K^0$, the resulting $s\bar s$ quark pair undergoes
hadronization with a $q\bar q$ pair from the vacuum, where $q\bar q \equiv u\bar u + d\bar d + s\bar s$, to form a meson-meson pair. The remaining $u$ quark combines with the initial $ud$ quark pair to form the proton $p$. This process can be written as

\begin{equation}
\begin{aligned}
|\Lambda_c\rangle &= \frac{1}{\sqrt{2}} |c(ud - du)\rangle  \\
 &\Rightarrow V^{(a)} s(\bar{u}u + \bar{d}d + \bar{s}s)\bar{s}u \frac{1}{\sqrt{2}}(ud - du) \\
&= V^{(a)} \left(M^2\right)_{33} p,
\end{aligned}
\end{equation}
where $V^{(a)}$ denotes the strength of the weak interaction, and the baryon wavefunctions $|p\rangle = \frac{1}{\sqrt{2}} |u(ud - du)\rangle$ and $|\Lambda_c\rangle = \frac{1}{\sqrt{2}} |c(ud - du)\rangle$ are used. $M$ denotes the $q\bar q$ matrix,
\begin{equation}
M=\begin{pmatrix}
u\bar u & u\bar d & u\bar s\\
d\bar u & d\bar d & d\bar s\\
s\bar u & s\bar d & s\bar s
\end{pmatrix}.
\label{eq:M}
\end{equation}
The matrix M, in terms of pseudoscalar mesons, is written as,
\begin{equation}
\resizebox{\columnwidth}{!}{$
M\Rightarrow P=
\begin{pmatrix}
\dfrac{\pi^0}{\sqrt2}+\dfrac{\eta}{\sqrt3}+\dfrac{\eta'}{\sqrt6} & \pi^+ & K^+\\[2mm]
\pi^- & -\dfrac{\pi^0}{\sqrt2}+\dfrac{\eta}{\sqrt3}+\dfrac{\eta'}{\sqrt6} & K^0\\[2mm]
K^- & \bar K^0 & -\dfrac{\eta}{\sqrt3}+\dfrac{2\eta'}{\sqrt6}
\end{pmatrix}$}.
\label{eq:P}
\end{equation}
Then, we have,
\begin{equation}
\begin{aligned}
\mathcal{Q}^{(a)}&= V^{(a)}(M^2)_{33}\,p \\
&= V_P V_{cs}V_{us}\Big(K^+K^- + K^0\bar K^0 + \frac{1}{3}\eta\eta\Big)p,
\end{aligned}
\label{eq:Ha-meson}
\end{equation}
where $V_P$ is the strength of the production vertex containing all the
dynamical factors~\cite{Wang:2020pem}, and $V_{cs}$, $V_{us}$ are the Cabibbo-Kobayashi-Maskawa (CKM) 
matrix elements. The $\eta'\eta'$ channel is neglected in
Eq.~\eqref{eq:Ha-meson} because the mass of $\eta'$ is too large.  We take $V_{us} = -0.22534$ and
$V_{cs} = 0.97427$~\cite{PDG2024}.

%图2
\begin{figure}[tbhp]
	\centering
	\includegraphics[scale=0.9]{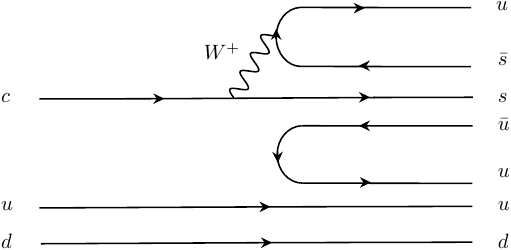}
	%\vspace{-0.7cm}
	\caption{Same as Fig.~\ref{fig:internalW}, but for the $W^+$ external-emission mechanism.}
	\label{fig:externalW}
\end{figure}

Additionally, the $\Lambda_c^+ \to p K^0 \bar{K}^0$ decay can proceed via the color-allowed external $W$ emission mechanism. At the quark level, the charm quark of the $\Lambda_c^+$ weakly decays into an $s$ quark and a $W^+$ boson, and the latter subsequently decays into a $u\bar{s}$ quark pair, which hadronizes into $K^+$. The remaining $s$ quark and the spectator $ud$ diquark combine with a $u\bar{u}$ quark pair created from the vacuum and hadronize into $K^-p$, as shown in Fig.~\ref{fig:externalW}. Then, we have
\begin{equation}
\begin{aligned}
\mathcal{Q}^{(b)} &= V^{(b)}(u\bar{s})s\bar{u}u \frac{1}{\sqrt{2}}(ud - du) \\
&= C \times V_P V_{cs} V_{us} K^+ K^- p,
\end{aligned}
\end{equation}
where we use the same normalization factor $V_p$ as in Eq.~\eqref{eq:Ha-meson}. The color factor $C$ represents the relative weight of the $W^+$ external-emission mechanism with respect to the $W^+$ internal-emission mechanism. In the external-emission diagram, the quark pair produced from the decay of the $W^+$ boson can have three possible color configurations, whereas in the internal-emission diagram, the quark colors are fixed. This color counting suggests that $C$ is theoretically expected to be 3. However, in actual weak decays of charmed baryons, $C$ may also be affected by the hadronization process, nonperturbative QCD effects, final-state interactions, and other related effects~\cite{Wang:2022nac}. We therefore do not fix $C$ a priori, but retain this factor in the subsequent theoretical derivation and determine its best-fit value in Sec.~\ref{sec:Results and Discussion} by comparing the resulting invariant mass squared distributions with the experimental data.

It should be noted that, in this work, the theoretical results are ultimately compared with the Belle measurement of the $K_S^0K_S^0$ invariant mass squared distribution. Therefore, the relation between $K^0\bar{K}^0$ and $K_S^0K_S^0$ must be taken into account. The neutral $K$ mesons can be expressed as
\begin{equation}
|K^0\rangle = \frac{1}{\sqrt{2}}|K_S^0\rangle + \frac{1}{\sqrt{2}}|K_L^0\rangle,
\end{equation}
\begin{equation}
|\bar{K}^0\rangle = \frac{1}{\sqrt{2}}|K_S^0\rangle - \frac{1}{\sqrt{2}}|K_L^0\rangle.
\label{eq:Kbar}
\end{equation}
The $K^0\bar K^0$ state can be expanded as

\begin{align}
|K^0\bar{K}^0\rangle=  &\frac{1}{2}|K_S^0K_S^0\rangle - \frac{1}{2}|K_L^0K_L^0\rangle \nonumber\\
&- \frac{1}{2}|K_S^0K_L^0\rangle + \frac{1}{2}|K_L^0K_S^0\rangle.
\end{align}
Consequently, projecting the $K^0\bar K^0$ state onto the $K_S^0K_S^0$ final state gives rise to a fixed factor of $1/2$ in the decay amplitude. Since $V_{cs}$, $V_{us}$ and the projection factor are all fixed quantities independent of the final-state invariant masses, these fixed factors are uniformly absorbed into the overall normalization factor at the production vertex in the subsequent fit to the Belle data, which we continue to denote by $V_P$~\cite{Dai:2018nmw,Zhang:2020rqr,Bayar:2014qha}. 

\begin{figure}[htbp]
	\centering
	\begin{minipage}{0.8\columnwidth}
		\centering
		\includegraphics[scale=0.9]{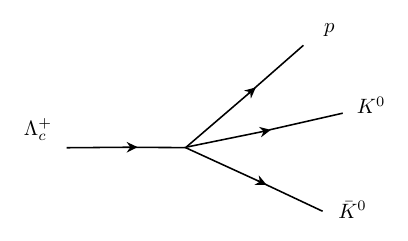}\\
		(a)
	\end{minipage}
	
	\vspace{0.5cm} % 两张图之间的垂直间距，可自行调整
	\begin{minipage}{0.8\columnwidth}
		\centering
		\includegraphics[scale=0.9]{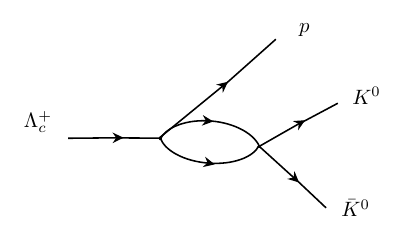}\\
		(b)
	\end{minipage}
	\caption{Hadron-level diagrams for the $\Lambda_c^+\to pK^0\bar K^0$ decay: (a) the tree diagram and (b) the rescattering mechanism.}\label{fig:tree and quan}
\label{fig:MB}
\end{figure}

After the meson pairs $K^+K^-$, $K^0\bar K^0$, and $\eta\eta$ are produced, the mesons undergo an $S$-wave final-state interaction. The corresponding hadron-level diagrams are shown in Fig.~\ref{fig:tree and quan}, where Fig.~\ref{fig:tree and quan}(a) is the tree diagram and Fig.~\ref{fig:tree and quan}(b) is the rescattering mechanism. Following this process, the amplitude for the $\Lambda_c^+\to pK_S^0 K_S^0$ decay in the charge basis is given by
\begin{equation}
t_{\Lambda_c^+\to pK_S^0 K_S^0}^{S\text{-wave}}
=V_p\left[h_{K^0\bar K^0}+\sum_i h_i G_i\,t_{i\to K^0\bar K^0}\right],
\label{eq:amp-charge}
\end{equation}
where $i=K^+K^-,K^0\bar K^0,\eta\eta$, and the weight factors are
\begin{equation}
h_{K^+K^-}=(1+C),\qquad h_{K^0\bar K^0}=1,\qquad h_{\eta\eta}=\frac13 .
\label{eq:h}
\end{equation}

However, the hadronically produced $K^+K^-$ and $K^0\bar{K}^0$ are states of definite charge rather than eigenstates of definite isospin. Both the $K$ and $\bar{K}$ are $I=1/2$ doublets, taken here as $(K^+,K^0)$ and $(\bar{K}^0,-K^-)$. Using the Clebsch--Gordan coefficients, the $K\bar{K}$ two-meson states in the charge basis can be decomposed into states of total isospin $I=0$ and $I=1$:
\begin{align}
|K^+K^-\rangle&=-\frac{1}{\sqrt2}|K\bar K\rangle^{I=0}
-\frac{1}{\sqrt2}|K\bar K\rangle^{I=1},\\
|K^0\bar K^0\rangle&=-\frac{1}{\sqrt2}|K\bar K\rangle^{I=0}
+\frac{1}{\sqrt2}|K\bar K\rangle^{I=1}.
\label{eq:CG2}
\end{align}
Accordingly, the transition amplitudes $t_{K^+K^-\to K^0\bar K^0}$ and $t_{K^0\bar K^0\to K^0\bar K^0}$ can be rewritten as
\begin{align}
t_{K^0\bar K^0\to K^0\bar K^0}
&=\frac12\,t_{K\bar K\to K\bar K}^{I=1}
+\frac12\,t_{K\bar K\to K\bar K}^{I=0},
\label{eq:t1}\\
t_{K^+K^-\to K^0\bar K^0}
&=-\frac12\,t_{K\bar K\to K\bar K}^{I=1}
+\frac12\,t_{K\bar K\to K\bar K}^{I=0}.
\label{eq:t2}
\end{align}
Taking $G_{K^0\bar K^0}=G_{K^+K^-}=G_{K\bar K}$ and combing with Eqs.~\eqref{eq:t1} and \eqref{eq:t2}, the amplitude of Eq.~\eqref{eq:amp-charge} becomes
\begin{eqnarray}
t_{\Lambda_c^+\to pK_S^0 K_S^0}^{S\text{-wave}}
&=& V_P\times h_{K^0\bar K^0} \nonumber \\
&& +V_P\Big\{\Big(1+\tfrac{C}{2}\Big)G_{K\bar K}
\big(t_{K^0\bar K^0 \to K^0\bar K^0} \nonumber \\
&& + t_{K^+K^- \to K^0\bar K^0}\big)+ \tfrac{1}{3}G_{\eta\eta}\,\tilde t_{\eta\eta \to K^0\bar K^0} \Big\} \nonumber\\
&& + V_P\Big\{ \Big(-\tfrac{C}{2}\Big)G_{K\bar K}
\big(t_{K^0\bar K^0 \to K^0\bar K^0}\nonumber\\
 &&- t_{K^+K^- \to K^0\bar K^0}\big) \Big\} \nonumber \\
&=& t^{\rm tree} + t^{I=0} + t^{I=1}.
\label{eq:amp_iso}
\end{eqnarray}
The amplitudes $ t^{\rm tree}$, $t^{I=0}$, and $t^{I=1}$ represent the contributions from the tree diagram, the $I=0$, and $I=1$ channels. Since $f_0(980)$ and $a_0(980)$ have isospin $I=0$ and $I=1$, respectively, the $I=0$ scattering amplitude is associated with the dynamical generation of $f_0(980)$, while the $I=1$ amplitude is associated with that of $a_0(980)$. For convenience, we denote $t^{I=0}$ and $t^{I=1}$ by $t^{f_0(980)}$ and $t^{a_0(980)}$ in the following.

In Eq.~(\ref{eq:amp_iso}), the combinatorial and symmetry factors associated with the identical $\eta$ mesons in the $\eta\eta$ intermediate state should be taken into account. The production of the $\eta\eta$ state involves two equivalent particle matchings, giving rise to a factor of $2$, while a symmetry factor of $1/2$ is included in the rescattering loop involving two identical particles \cite{Wang:2020pem}. The scattering matrix $t_{i\to j}$ has been calculated within the chiral unitary approach in Refs. \cite{Dias:2016gou,Oller:1997ti,Oller:2000ma}.
For the $\eta\eta$ intermediate state, we adopt the normalization convention for identical particles \cite{Aceti:2012dd,Ikeno:2021kzf,Dias:2016gou}, and take 
$\tilde{t}_{\eta\eta\to j} = \sqrt{2}\,t_{\eta\eta\to j}$. Combined with the
factor of 2 from the two equivalent matchings and the symmetry factor $1/2$ of
the loop, this normalization is chosen so that the resulting $\eta\eta$
contribution is fully consistent with the standard phase-space normalization
for two identical particles used in Ref.~\cite{Wang:2020pem}.

$G_l$ in Eq.~(\ref{eq:amp_iso}) denotes the two-meson loop function for the $l$-th channel. After performing the ${d}q^0$ integration, the loop function can be written as~\cite{Oller:2000ma},
\begin{equation}
\begin{aligned}
G_l&=i\int\frac{d^4q}{(2\pi)^4}
 \frac{1}{(p-q)^2-m_1^2+i\epsilon}\,
 \frac{1}{q^2-m_2^2+i\epsilon}\\
&=\int\frac{d^3\mathbf q}{(2\pi)^3}
 \frac{\omega_1+\omega_2}{\omega_1\omega_2}
 \frac{1}{(\sqrt s+\omega_1+\omega_2)
 (\sqrt s-\omega_1-\omega_2+i\epsilon)} ,
\end{aligned}
\label{eq:G}
\end{equation}
where $\sqrt{s}$ denotes the invariant mass of the meson-meson system and $\omega_i$ is the energy of the $i$-th meson, 
\begin{equation}
\omega_i = \sqrt{q^2 + m_i^2}, \quad i=1,2.
\end{equation}
The three-momentum integral in Eq.~\eqref{eq:G} is regularized by introducing a momentum cutoff $|\mathbf q_{\max}|=600~\mathrm{MeV}$,
consistent with the value adopted in Refs.~\cite{Wang2021a0f0,Wang:2020pem,Lyu:2025oow,Duan:2024czu}. The transition amplitude $t_{i\to j}$ is obtained by solving the Bethe-Salpeter equation in the coupled-channel space~\cite{Oller:1997ti},
\begin{equation}
T=[1-VG]^{-1}V.
\label{eq:BS}
\end{equation}
For $I=0$, five coupled channels, $\pi^+\pi^-$, $\pi^0\pi^0$, $K^+K^-$, $K^0\bar K^0$, and $\eta\eta$ are included, while for $I=1$, three channels $K^+K^-$, $K^0\bar K^0$, and $\pi^0\eta$ are considered. The matrix $G$ is diagonal, with its diagonal elements given by the loop function defined in Eq.~\eqref{eq:G}. The matrix $V$ denotes the interaction kernel, whose elements correspond to the tree-level transition amplitudes derived from the phenomenological Lagrangian in Refs.~\cite{Oller:1997ng,Oller:1997ti}. For the $I=0$ sector, the explicit expressions of $V$ can be written as~\cite{Wang:2020pem,Zhang:2024myn},
\begin{align}
&V_{\pi^+\pi^-\to\pi^+\pi^-}=-\frac{1}{2f^2}s, \notag\\
&V_{\pi^+\pi^-\to\pi^0\pi^0}=-\frac{1}{\sqrt2 f^2}(s-m_\pi^2), \notag\\
&V_{\pi^+\pi^-\to K^+K^-}=-\frac{1}{4f^2}s, \notag\\
&V_{\pi^+\pi^-\to K^0\bar K^0}=-\frac{1}{4f^2}s, \notag\\
&V_{\pi^+\pi^-\to\eta\eta}=-\frac{1}{3\sqrt2 f^2}m_\pi^2, \notag\\
&V_{\pi^0\pi^0\to\pi^0\pi^0}=-\frac{1}{2f^2}m_\pi^2, \notag\\
&V_{\pi^0\pi^0\to K^+K^-}=-\frac{1}{4\sqrt2 f^2}s, \notag\\
&V_{\pi^0\pi^0\to K^0\bar K^0}=-\frac{1}{4\sqrt2 f^2}s, \notag\\
&V_{\pi^0\pi^0\to\eta\eta}=-\frac{1}{6f^2}m_\pi^2, \notag\\
&V_{K^+K^-\to K^+K^-}=-\frac{1}{2f^2}s, \notag\\
&V_{K^+K^-\to K^0\bar K^0}=-\frac{1}{4f^2}s, \notag\\
&V_{K^+K^-\to\eta\eta}=-\frac{1}{12\sqrt2 f^2}(9s-6m_\eta^2-2m_\pi^2), \notag\\
&V_{K^0\bar K^0\to K^0\bar K^0}=-\frac{1}{2f^2}s, \notag\\
&V_{K^0\bar K^0\to\eta\eta}=-\frac{1}{12\sqrt2 f^2}(9s-6m_\eta^2-2m_\pi^2), \notag\\
&V_{\eta\eta\to\eta\eta}=-\frac{1}{18f^2}(16m_K^2-7m_\pi^2). \label{eq:V_final}
\end{align}
For the $I=1$ sector the potentials are~\cite{Wang:2020pem},
\begin{align}
&V_{K^+K^-\to K^+K^-}=-\frac{1}{2f^2}s, \notag\\
&V_{K^+K^-\to K^0\bar K^0}=-\frac{1}{4f^2}s, \notag\\
&V_{K^0\bar K^0\to K^0\bar K^0}=-\frac{1}{2f^2}s, \notag\\
&V_{\pi^0\eta\to\pi^0\eta}=-\frac{m_\pi^2}{3f^2}, \notag\\
&V_{K^+K^-\to\pi^0\eta}=-\frac{\sqrt{3}}{12f^2}\left(3s-\frac{8}{3}m_K^2-\frac{1}{3}m_\pi^2-m_\eta^2\right), \notag\\
&V_{K^0\bar K^0\to\pi^0\eta}=-V_{K^+K^-\to\pi^0\eta}. \label{eq:V_final_I1}
\end{align}
Here $f=f_\pi=93~\mathrm{MeV}$ is the pion decay constant, $s$ is the invariant mass squared of the meson pair, and $m_\pi$, $m_K$, and $m_\eta$ are the averaged masses of the pion, kaon, and $\eta$ mesons, respectively~\cite{PDG2024}. Solving Eq.~\eqref{eq:BS} gives the transition amplitudes $t_{i\to K^0\bar K^0}$ from the various coupled channels to the $K^0\bar K^0$ channel.
We note that the $I=1$ amplitude used here incorporates not only the
$K\bar K$ channel but also the open $\pi\eta$ channel. This is essential for a
faithful description of the $a_0(980)$ line shape near threshold, and cannot be
captured by a single-channel Flatt\'e parametrization.
%------------------------------------------------------------------
\subsection{Contribution from the $\Sigma(1620)$}
\label{sec2b}
%------------------------------------------------------------------
\begin{figure}[tbhp]
 	\centering
 	\includegraphics[scale=0.9]{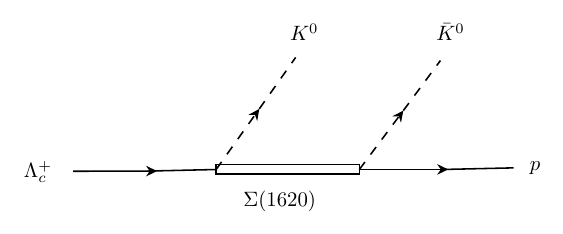}
 	%\vspace{-0.7cm}
 	\caption{Hadronic-level diagram for the $\Lambda_c^+ \to p K^0\bar K^0$ decay process, considering the contributions of the intermediate resonance $\Sigma(1620)$.}
 	\label{fig:1620}
 \end{figure}

As discussed in the Introduction, the Belle collaboration reported mild excesses in the \(pK_S^0\) invariant mass squared distribution around \(M^2_{pK_S^0}\approx 2.6\text{--}2.8~\mathrm{GeV}^2\). These features lie close to the mass of the strange baryon resonance \(\Sigma(1620)\)~\cite{PDG2024}. We therefore consider a possible \(\Sigma(1620)\) contribution in the \(p\bar K^0\) final state.

Before proceeding, we note two constraints imposed by charge conservation.
Since the $pK^0$ and $p\bar K^0$ subsystems both carry charge $+1$ while the
$\Lambda$ excitations are neutral, no $\Lambda$ excitation can contribute to
the $pK_S^0$ spectrum. The only allowed intermediate baryon resonance is
therefore a $\Sigma$-type excitation, which contains a charged $Q=+1$ member.
In the following we retain $\Sigma(1620)$ as the representative candidate and
examine whether the data require it.

The corresponding mechanism is shown in Fig.~\ref{fig:1620}, where the
intermediate resonance $\Sigma(1620)$ decays into the $p\bar K^0$ final state. We describe the
resonant amplitude by the Breit-Wigner form
\begin{equation}
t^{\Sigma(1620)} = \frac{V_{\Sigma(1620)} M_{\Sigma(1620)} \Gamma_{\Sigma(1620)}}
                       {M_{p\bar K^0}^2 - M_{\Sigma(1620)}^2 + i M_{\Sigma(1620)} \Gamma_{\Sigma(1620)}},\label{eq:T_Sigma1620}
\end{equation}
with $M_{\Sigma(1620)} = 1620$~MeV and $\Gamma_{\Sigma(1620)} = 70$~MeV, taken
from the  Review of Particle Physics (RPP)~\cite{PDG2024}. Here $V_{\Sigma(1620)}$ denotes the relative weight of the
intermediate $\Sigma(1620)$, which is determined by fitting the Belle
data~\cite{Belle:2022pwd}.

%------------------------------------------------------------------
\subsection{Invariant mass distributions}
\label{sec2c}
%------------------------------------------------------------------
The contributions of the $\Sigma(1620)$ in the present process have not yet been confirmed experimentally. To test whether such an intermediate resonance appears in the $\Lambda_c^+\to pK_S^0K_S^0$ decay, we first consider the case in which it is absent, namely the amplitude from the $S$-wave meson-meson interaction of Sec.~\ref{sec2a}. This is denoted as ``Model~A'':
\begin{equation}
|T^A|^2=\left|t^{\rm tree}+t^{f_0(980)}+t^{a_0(980)}e^{i\phi_1}\right|^2,
\label{eq:modelA}
\end{equation}
where $\phi_1$ is the relative phase between the $f_0(980)$ and $a_0(980)$.
Based on the Model~A, we add the $\Sigma(1620)$ Breit--Wigner amplitude of Sec.~\ref{sec2b}, giving ``Model~B''
\begin{align}
|T^B|^2&=\left|t^{\rm tree}+t^{f_0(980)} + t^{a_0(980)} e^{i\phi_1}+t^{\Sigma(1620)}e^{i\phi_2}\right|^2,
\label{eq:modelB}
\end{align}
where $\phi_2$ is the relative phases of the $\Sigma(1620)$ with respect to the $f_0(980)$.

The double-differential decay width of the $\Lambda_c^+\to pK^0\bar K^0$ process is then
\begin{equation}
\frac{d^2\Gamma}{dM_{p\bar K^0}^2\,dM_{K^0\bar K^0}^2}
=\frac{1}{(2\pi)^3}\frac{1}{32M_{\Lambda_c^+}^3}\,|T|^2 ,
\label{eq:d2Gamma1}
\end{equation}
\begin{equation}
\frac{d^2\Gamma}{dM_{pK^0}^2\,dM_{K^0\bar K^0}^2}
=\frac{1}{(2\pi)^3}\frac{1}{32M_{\Lambda_c^+}^3}\,|T|^2 ,
\label{eq:d2Gamma2}
\end{equation}
where $|T|^2$ takes the values $|T^A|^2$ and $|T^B|^2$ in the two models.

%In the theoretical calculation, the $S$-wave final-state interaction is included in the $K^0\bar K^0$ system, while the intermediate resonances are coupled to the $p\bar K^0$ system.

Because the two $K_S^0$ mesons are experimentally indistinguishable, both
possible combinations of the proton $p$ with the two $K_S^0$ mesons are
included. Therefore the $pK_S^0$ invariant mass squared distribution is given by the
sum of the projections onto the $pK^0$ and $p\bar K^0$ subsystems,
\begin{equation}
\frac{d\Gamma}{dM_{pK_S^0}^2}
=\frac{d\Gamma}{dM_{pK^0}^2}+\frac{d\Gamma}{dM_{p\bar K^0}^2}.
\label{eq:proj}
\end{equation}

By integrating Eqs.~\eqref{eq:d2Gamma1} and \eqref{eq:d2Gamma2}, we obtain the invariant mass squared distributions $d\Gamma/dM_{pK^0}^2$, $d\Gamma/dM^2_{p\bar K^0}$, and $d\Gamma/dM^2_{K^0\bar K^0}$.   Given a specific value of the invariant mass squared $M_{12}^2$, the range of $M_{23}^2$ is determined by~\cite{PDG2024},
\begin{align}
	&\left(m_{23}^2\right)_{\min}=\left(E_2^*+E_3^*\right)^2-\left(\sqrt{E_2^{* 2}-m_2^2}+\sqrt{E_3^{* 2}-m_3^2}\right)^2, \nonumber\\
	&\left(m_{23}^2\right)_{\max}=\left(E_2^*+E_3^*\right)^2-\left(\sqrt{E_2^{* 2}-m_2^2}-\sqrt{E_3^{* 2}-m_3^2}\right)^2, \label{eq:limit}
\end{align}
where $E_2^*$ and $E_3^*$ are the energies of final particles 2 and 3 in the $M_{12}$ rest frame, respectively, which are written as,
\begin{align}
E_2^*&=\frac{M_{12}^2-m_1^2+m_2^2}{2M_{12}},
\label{eq:E2}\\
E_3^*&=\frac{M_{\Lambda_c^+}^2-M_{12}^2-m_3^2}{2M_{12}},
\label{eq:E3}
\end{align}
with $m_1$, $m_2$, and $m_3$ are the masses of involved particles 1, 2, and 3, respectively. The particle masses and widths are taken from the RPP~\cite{PDG2024}.

%%%%%%%%%%%%%%%%%%%%%%

\section{Results and Discussion} \label{sec:Results and Discussion}

%kk振幅表示图
%\begin{figure}[tbhp]
%    \centering
%    \hspace*{-0.45cm}\includegraphics[scale=0.9]{figures/amplitude_KK.pdf}
%    \caption{}
%    \label{fig:amplitude_KK}
%\end{figure}

\begin{table}[b]
\caption{Fitted parameters and $\chi^2/\text{d.o.f.}$ for Model~A and
Model~B.}
\label{tab1}
\begin{ruledtabular}
\begin{tabular}{lcc}
Parameter & Model~A & Model~B \\
\hline
$C$ & $2.914$ & $2.507$ \\
$V_P$ & $3892.4$ & $3881.0$ \\
$V_{\Sigma(1620)}$ & --- & $2731.6$ \\
$\phi_1$ & $0.723\pi$ & $0.796\pi$ \\
$\phi_2$ & --- & $1.226\pi$ \\
$\chi^2/\text{d.o.f.}$ & $2.287$ & $2.275$ \\
\end{tabular}
\end{ruledtabular}
\end{table}

\begin{figure*}[htbp]
	\centering
	\includegraphics[width=0.48\textwidth]{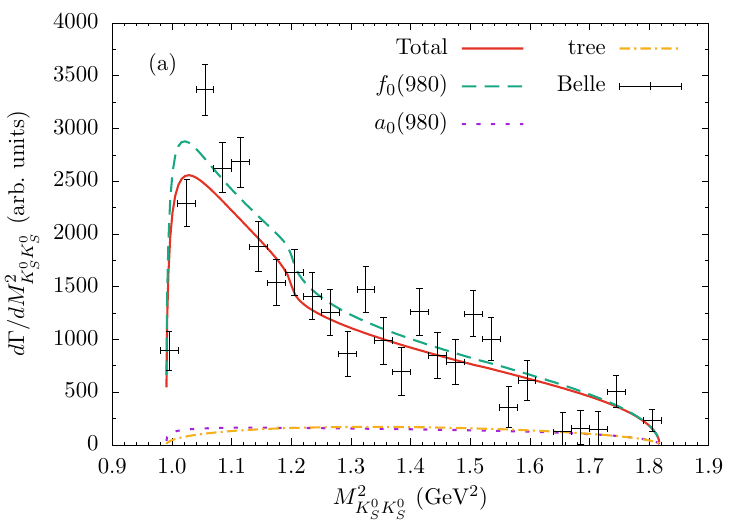}\hfill
	\includegraphics[width=0.48\textwidth]{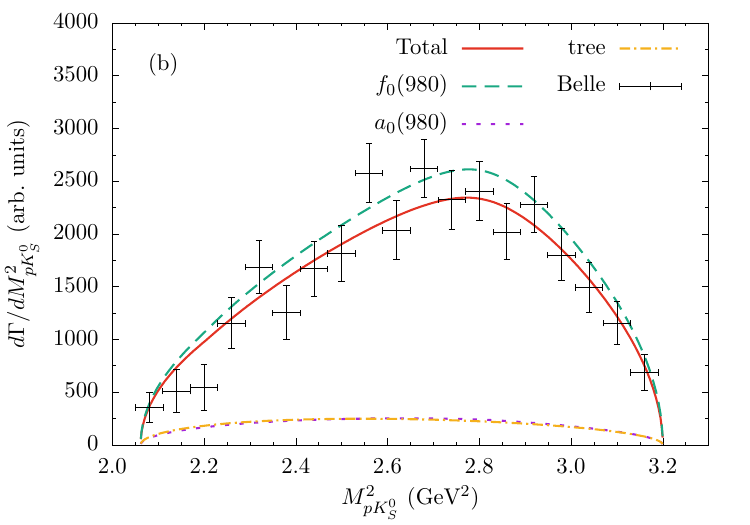}
	\par\vspace{0.3cm}
	\includegraphics[width=0.48\textwidth]{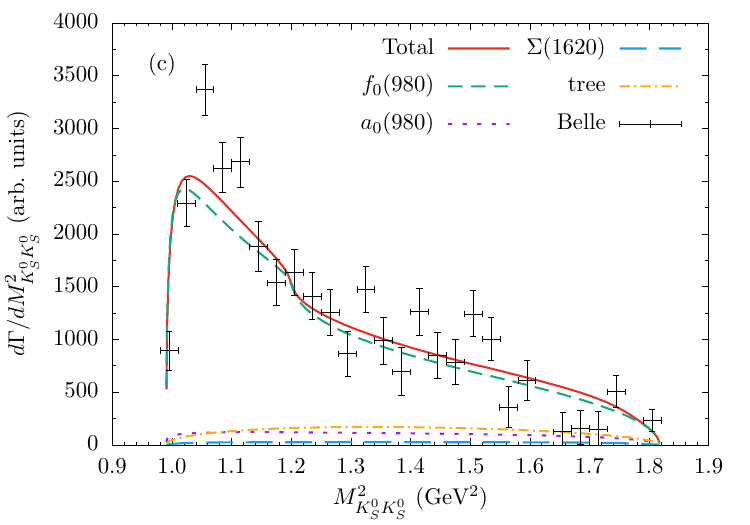}\hfill
	\includegraphics[width=0.48\textwidth]{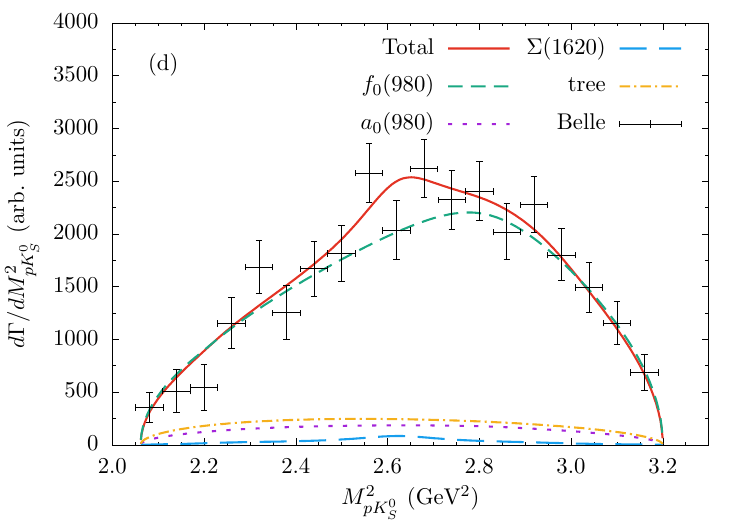}
	\caption{$K_S^0K_S^0$ and $pK_S^0$ invariant mass squared distributions of
the process $\Lambda_c^+ \to pK_S^0K_S^0$ with the fitted parameters of
Model~A (a,b) and Model~B (c,d). The black data points with error bars labeled
``Belle'' are the Belle data taken from Ref.~\cite{Belle:2022pwd}, and the solid
curves labeled ``Total'' show our theoretical results for the total
contributions. The contributions from the $f_0(980)$, $a_0(980)$, 
$\Sigma(1620)$, and tree-level are also shown, labeled ``$f_0(980)$'', ``$a_0(980)$'', 
``$\Sigma(1620)$'', and ``tree'', respectively.}
	\label{fig:result}
\end{figure*}

%Dalitz图-KSKS

 \begin{figure}[tbhp]
    \centering
    \hspace{-0.6cm}\includegraphics[scale=0.6]{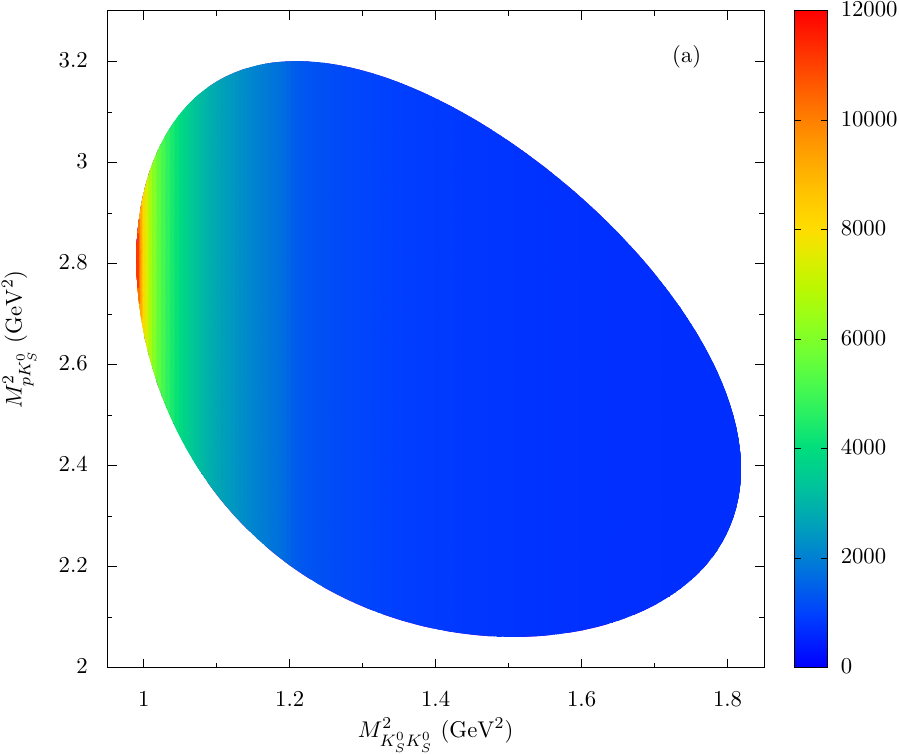}

    \vspace{0.4cm}
    \hspace{-0.6cm}\includegraphics[scale=0.6]{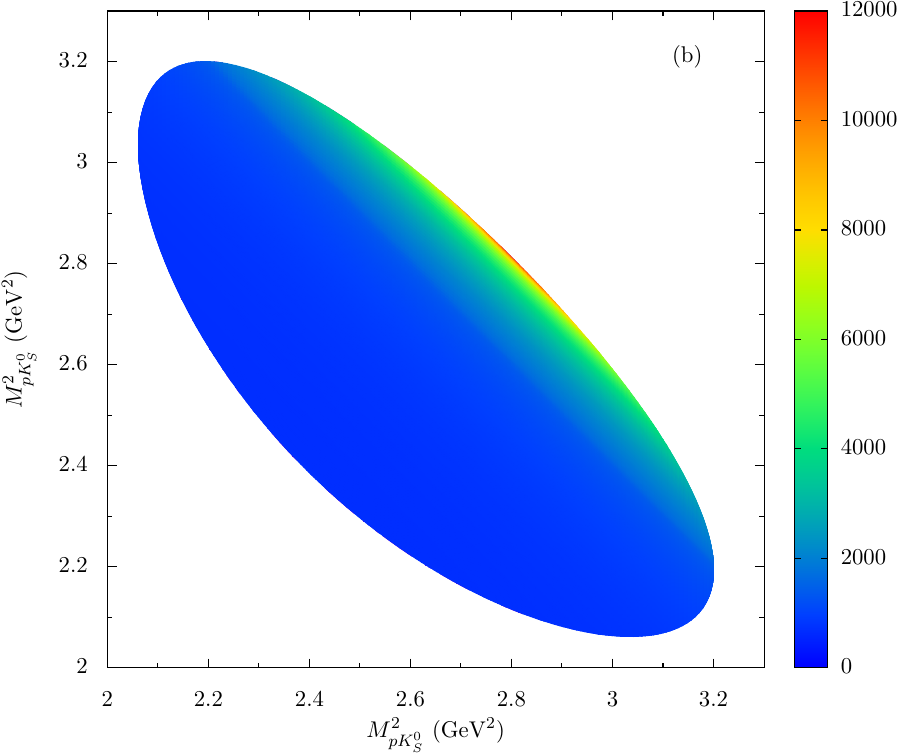}
    \caption{Dalitz plot of the process
$\Lambda_c^+ \to p K_S^0 K_S^0$ with the fitted
parameters of Model A.}
    \label{fig:DalitzA-KsKs}
\end{figure}

The theoretical framework contains five free parameters: (1) the color factor $C$; (2) the overall normalization $V_P$ of the $S$-wave meson-meson amplitude; (3) the relative weight $V_{\Sigma(1620)}$ of the intermediate baryon resonances; and (4) the interference phases $\phi_1$ and $\phi_2$.
We fit Model~A and Model~B to the
$K_S^0 K_S^0$ and $p K_S^0$ invariant mass squared distributions measured by the Belle Collaboration~\cite{Belle:2022pwd}, which have 45 effective data points in total, and are efficiency-corrected and background-subtracted~\cite{Longke}. The fitted parameters and the value of $\chi^2/\mathrm{d.o.f.}$ are summarized in Table~\ref{tab1}.

The $\chi^2/\mathrm{d.o.f.}$ obtained with Model A is 2.287, and the fitted color factor $C = 2.914$ is in agreement with the theoretical expectation $C\sim 3$. 
The invariant mass squared distributions of the $\Lambda_c^+\to pK_S^0K_S^0$ decay, calculated with the fitted parameters in Table~\ref{tab1}, are shown in Fig.~\ref{fig:result}, where panel~(a) is the $K_S^0 K_S^0$ spectrum corresponding to the Belle $K_S^0K_S^0$ spectrum, and panel~(b) is the $pK_S^0$ spectrum given by the sum of the $pK^0$ and $p\bar K^0$ projections. 

The $S$-wave $K\bar{K}$ final-state interaction alone already describes fairly well the enhancement of the $K_S^0K_S^0$ spectrum near threshold, showing that the $K\bar K$ final-state interaction plays an important dynamical role in this process. The $f_0(980)$ is the main source of the near-threshold structure, while the $a_0(980)$ contribution  is relatively small.
The same $I=0$ dominance
has been obtained in the charmless $B^- \to \pi^-K^+K^-$ and
$B^- \to \pi^-K^0\bar K^0$ reactions within the chiral unitary SU(3) formalism,
where the $I=0$ component associated with the $f_0(980)$ was found to generate
the dominant contribution in the low $K^+K^-$ invariant mass
region~\cite{Abreu:2023hts}. Our result shows that this pattern also holds in the
charmed-baryon sector, and that it survives even in a channel that is free of
the $\phi$ peak.

To further investigate the possible intermediate baryon-resonance structures in the $pK_S^0$ invariant mass distribution, we introduce $\Sigma(1620)$ into Model A. 
The fitted parameters are summarized in Table~\ref{tab1}.
The inclusion of $\Sigma(1620)$ leads to changes in the $pK_S^0$ line shape, but
the fit quality has no sizeable improvement: the $\chi^2/\text{d.o.f.}$ changes from 2.287
(Model~A) to 2.275 (Model~B). 
It implies that the present data dose not favor a sizeable $\Sigma(1620)$ contribution.
This is consistent with the expectation that the $\Sigma$-type background in
this channel is weak, and it underlines the conclusion that the $pK_S^0$
spectrum is dominated by the $K\bar K$ final-state interaction rather than by
baryon-resonance dynamics.

In addition, as shown in Fig.~\ref{fig:DalitzA-KsKs}, we present the Dalitz plots of $M^2_{K_S^0K_S^0}$ versus $M^2_{pK_S^0}$ and $M^2_{pK_S^0}$ versus $M^2_{pK_S^0}$  for the $\Lambda_c^+ \to p K_S^0K_S^0$ decay using the fitted parameters of Model A. One can clearly observe the signal of the resonance $f_0(980)$.

Before closing this section, we comment on the description of the $f_0(980)$
and $a_0(980)$ line shapes near the $K\bar K$ threshold. In many analyses these
two states are parametrized by the Flatt\'e form, in which only the coupling to
the $K\bar K$ channel is retained and the other coupled channels are neglected.
Such a treatment is adequate for reproducing the gross shape of a
near-threshold enhancement, but it is incomplete for the $a_0(980)$: this state
has $I=1$ and couples significantly to the $\pi\eta$ channel, which is open in
the relevant energy region and whose threshold lies close to that of $K\bar K$.
Neglecting the $\pi\eta$ channel distorts both the position and the width
extracted for $a_0(980)$, and, through the $I=0$ and $I=1$ interference that
controls the line shape of the $K_S^0K_S^0$ spectrum, it also affects the
extracted $f_0(980)$ parameters. In the present work the $f_0(980)$ and
$a_0(980)$ are generated from the $S$-wave pseudoscalar meson--pseudoscalar
meson interaction within the chiral unitary approach: the amplitude is obtained
by solving the coupled-channel Bethe-Salpeter equation, Eq.~\eqref{eq:BS}, with
the $I=0$ ($\pi\pi$, $K\bar K$, $\eta\eta$) and $I=1$ ($K\bar K$, $\pi\eta$)
channels treated on the same footing. The resulting amplitudes incorporate the
full coupled-channel dynamics and the correct analytic structure near
threshold, and therefore provide a more faithful and internally consistent
description of the dynamical properties of $f_0(980)$ and $a_0(980)$ than a
single-channel Flatt\'e parametrization. A similar philosophy has proved
essential in the amplitude analysis of $J/\psi \to \gamma K_S^0K_S^0\pi^0$,
where the description of the pseudoscalar component in the $K_S^0K_S^0\pi^0$
invariant mass distribution required a consistent treatment of the resonance
dynamics and of the overlapping structures in the
$K_S^0K_S^0\pi^0$ system~\cite{BESIII:2022chl}.

%==================================================================
\section{Summary}
\label{sec:summary}
%==================================================================
Recently, the Belle Collaboration measured the $\Lambda_c^+ \to pK_S^0K_S^0$ decay, obtaining a branching fraction of $(2.35 \pm 0.12 \pm 0.07 \pm 0.12)\times 10^{-4}$ and providing the invariant mass squared distributions of $K_S^0K_S^0$ and $pK_S^0$~\cite{Belle:2022pwd}. A significant enhancement structure was found near the $K\bar{K}$ threshold in the $K_S^0K_S^0$ invariant mass spectrum.

In this work we have performed a theoretical study of $\Lambda_c^+ \to pK^0\bar K^0$ within the chiral unitary approach, based on the Belle data. We have argued that, compared with the charged channel $\Lambda_c^+ \to pK^+K^-$~\cite{Zhang:2026oml}, this channel is a cleaner probe of the $K\bar K$ final-state interaction: the $\phi(1020)$ contribution is forbidden by the antisymmetric-to-symmetric
projection, and, because both $pK^0$ and $p\bar K^0$ carry charge $+1$ while the $\Lambda$ excitations are neutral, no $\Lambda$ excitation can contribute. The only possible baryon-resonance background is a $\Sigma$-type excitation, and we
find that including $\Sigma(1620)$ does not improve the fit ($\chi^2/\text{d.o.f.}$ changing from 2.287 to 2.275).

The near-threshold enhancement is well described by the $S$-wave $K^0\bar K^0$ final-state interaction alone. The $f_0(980)$ contribution is dominant while that of $a_0(980)$ is relatively small, and the fitted color factor $C = 2.914$ agrees with the theoretical expectation $C \sim 3$. The same $I=0$ dominance has been observed in the low-$K^+K^-$ region of charmless $B$
decays~\cite{Abreu:2023hts}, and our analysis shows that it is a robust feature of
the $K\bar K$ final-state interaction rather than an artifact of a particular production mechanism or of the presence of a $\phi$ peak.

We have further emphasized that the near-threshold $f_0(980)$ and $a_0(980)$
are usually described by Flatt\'e parametrizations retaining only the $K\bar K$
coupling; the chiral-unitary amplitudes, in which the coupled-channel
dynamics---in particular the $\pi\eta$ channel for $a_0(980)$---is treated
consistently, provide a more faithful description of the dynamical structure of
these states.

Finally, we note that the spin structure of three-body charmed-baryon decays
opens a further, so far unexplored, direction. Because the three-body final
state in $\Lambda_c^+ \to pK_S^0K_S^0$ contains a spin-$\tfrac12$ baryon and a
pair of pseudoscalar mesons, the decay amplitude can be decomposed into
parity-conserving and parity-violating parts, and the resulting up-down
asymmetry of the final-state baryon provides a new observable that is sensitive
to the underlying weak and strong dynamics. In the $SU(3)_f$ framework, such
up-down asymmetries have been evaluated systematically for the three-body
anti-triplet charmed-baryon decays $B_c \to B_n MM'$, with the $s$-wave
meson-pair configurations taken as the dominant constituent of the final
state~\cite{Cen:2019ims}. Once a partial-wave or amplitude analysis of
$\Lambda_c^+ \to pK_S^0K_S^0$ becomes available, combining the measured up-down
asymmetry with the $K\bar K$ final-state interaction studied here would allow
one to probe the spin structure of the decay and to test the flavor symmetry of
the weak Hamiltonian, thereby extending the scope of the present analysis
beyond the spectroscopy of the light scalars.

Future measurements with higher precision, in particular from Belle~II and the
proposed Super Tau-Charm Facility~\cite{Achasov:2023gey}, will help test the roles of the $f_0(980)$
and $a_0(980)$ and of the possible $\Sigma(1620)$ contribution in this process.

\section*{Acknowledgments}
This work is supported by the National Key R$\&$D Program of China (No. 2024YFE0105200), the National Natural Science Foundation of China under Grant No. 12475086, No. 12192263.

%图5：拟合图8个

\bibliographystyle{unsrt}%规定参考文献的样式
%\bibliography{cite}  %参考文献库的名字Ref

\bibliography{ref}
\end{document}